\documentclass{iau}

\usepackage{amsmath}
\usepackage{graphicx}
\usepackage{multirow}
\usepackage{natbib}  % remove later?

\newcommand\farcs{\mbox{$.\!\!^{\prime\prime}$}}

\begin{document}

\lefttitle{T.~Joseph~W.~Lazio}
\righttitle{Solar System Technosignatures}

\jnlPage{}{}
\jnlDoiYr{2026}
\doival{10.1017/xxxxx}

\aopheadtitle{Proceedings IAU Symposium 404: Advancing the Search for Technosignatures}
\editors{J.~Haqq-Misra \& R.~Kopparapu, eds.}

\title{Technosignatures in the Solar System}

\author{T.~Joseph~W.~Lazio}
\affiliation{ }

\begin{abstract}
NASA has five robotic space probes on escape trajectories from the Solar System, and the Interstellar Probe concept was considered in the recent U.{}S.\ Solar \& Space Physics Decadal Survey.  While none of these robotic probes will be operational when they reach another star, it is natural to ask whether another civilization also might have sent out interstellar probes.  Serious consideration of interstellar probes dates at least to R.~Bracewell in the early 1960s, and the discovery of three interstellar objects has rekindled some of that interest.  I consider current limits on signatures of extraterrestrial technology in the Solar System, both objects on various orbits (``probes'') and surface artifacts, using data from planetary exploration and astronomical sky surveys.  Perhaps not surprisingly, the completeness to which the Solar System has been searched varies as a function of distance from the Sun.
I conclude that only extremely crude upper limits can be placed on the existence of technosignatures in the Solar System and that, in some cases, relatively large probes or surface artifacts would have escaped detection.
I also highlight areas that might be profitable for improving these limits considerably.
\end{abstract}

\begin{keywords}
Deep space probes,
%UAT 366
Lunar surface,
%https://astrothesaurus.org/uat/974
Planetary surfaces,
%https://astrothesaurus.org/uat/2113
Search for Extraterrestrial Intelligence,
%https://astrothesaurus.org/uat/2127
Space probes,
%https://astrothesaurus.org/uat/1545
Technosignatures
%https://astrothesaurus.org/uat/2128
\end{keywords}

\maketitle

\section{Introduction}\label{sec:intro}

The concept of interstellar probes constructed by other civilizations and present in the Solar System was introduced by \cite{b60}, if not earlier, and it continues to engender discussion~\citep{2016JBIS...69...88G,2019AJ....158..150B,2021JBIS...74...76B,2021SerAJ.203...47O,2025arXiv250816825D}.
The utility of artifacts as technosignatures, be they intentional carriers of a message or merely relics of exploration, has long been recognized \citep{b60}, because of their persistence (``time capsule'' or ``message in a bottle'' nature) and time-integrated data volume.
By analogy to the Search for Extraterrestrial Intelligence (SETI),
Freitas and collaborators proposed that the possibility of physical
technological objects should motivate a Search for Extraterrestrial
Artifacts
\citep[\hbox{SETA},][]{1983JBIS...36..501F,1985AcAau..12.1027F}.

The notion of physical technology objects from another civilization 
present in the Solar System is plausible based on the Pioneer~10 and~11, Voyager~1 and~2, and New Horizons spacecraft, all of which either have left the Solar System or are on escape trajectories.
Further, there have been multiple concept studies for missions
designed to leave the Solar System, the most recent of which was the
Interstellar Probe mission concept\footnote{
The Interstellar Probe concept was not recommended to proceed in this
decade in \textit{The Next Decade of Discovery in Solar and Space Physics: Exploring and Safeguarding Humanity's Home in Space} \citep{Heliophysics2024}.}
\citep{interstellarprobe}.
None of these spacecraft will be operational on the time scales
relevant for reaching a nearby star ($> 10\,000$\,yr), nor would the
Interstellar Probe concept would have been designed to be
operational on these time scales.
Nonetheless, the existence of these spacecraft establishes an existence proof that it is feasible to launch interstellar probes.

Even for passive probes, the total amount of information stored on such objects would be limited only by the total mass and available media used to store and retrieve data. From etched drawings to gold-plated data disks, many forms of hard-coded media could be used to ensure the persistence of information over time, impervious to the deleterious effects of Galactic cosmic rays. 
Indeed, \cite{rose2004inscribed} argued that the efficiency of inscribed matter for information storage on spacecraft sent from one civilization to another made them vastly more efficient for interstellar communications than transmissions of photons.

Whether operational or not, and whether aimed intentionally at the Solar System or not, finding a spacecraft or other artifact of non-terrestrial origin in the Solar System would be an unambiguous technosignature, if one could be identified.

\section{The Solar System Technosignature Matrix}\label{sec:matrix}

Table~\ref{tab:probes} is adapted from the W.~M.~Keck Institute for Space Studies report \textit{Data-Driven Approaches to Searches for the Technosignatures of Advanced Civilizations} \citep{KISS} and summarizes the classes of physical artifacts that might exist within the Solar System.
The categories shown are not exclusive in the sense that a physical object could transition from one class to another (e.g., from passive to active by charging batteries or from active to passive from a component failure).
It does, however, provide a means for structuring searches for physical technosignatures in the Solar System.

\begin{table}
\centering
    \caption{Classes of Physical Technosignatures\label{tab:probes}}
    \begin{tabular}{p{0.1\textwidth}|p{0.4\textwidth}|p{0.4\textwidth}}
    \noalign{\hrule\hrule}
                   & \textbf{Active} & \textbf{Passive} \\
    \noalign{\hrule}
    \textbf{Probe} & Object on orbit, including  hyperbolic, that uses an energy source, either internal or solar radiation, to conduct measurements or transmit signals, not necessarily aimed at Earth
          & Object on orbit, including hyperbolic, that undertakes no actions \\
    & & \\
    \noalign{\hrule}
    \textbf{Surface Artifact} & Object on surface of planetary body that uses an energy source, internal to it, from the body, or from solar radiation, to conduct measurements or transmit signals, not necessarily aimed at Earth
          & Object on surface of planetary body that undertakes no actions \\
    & & \\
    \noalign{\hrule}
    \end{tabular}
\end{table}

I now pose the following hypothesis:
\begin{center}
\parbox{0.95\textwidth}{%
    \textbf{Hypothesis}: One or more physical ET technosignatures are present in the Solar System today.}
\end{center}
In the next sections, I consider whether it is possible to falsify this hypothesis for each of the four classes of physical technosignatures defined in Table~\ref{tab:probes}.
My assessment will consider a variety of different possible investigations, though I may not consider every possible observation and there may be other observations discussed in the literature that I do not address here.
Throughout, I often will use the term ``probe'' as the more general term for an ET technological object.

A recurring theme in these assessments will be the difference between \emph{detecting} an ET technological object and \emph{identifying} it as an ET technology.

\subsection{Passive Probes}\label{sec:matrix.pprobe}

Identifying a passive probe is exactly the scenario that may apply in the future for any of humanity's current probes (Pioneer~10 or~11, Voyager~1 or~2, New Horizons) should or when one of them passes close to another star: It will be an inert object traversing that stellar system.

\begin{figure}
  \centering
  \includegraphics[width=0.97\textwidth]
    %alt="An image from a telescope showing a central dot, which is reflected light from an interstellar comet, and stars, which appear as streaks because the telescope was tracking the interstellar comet."]%
  {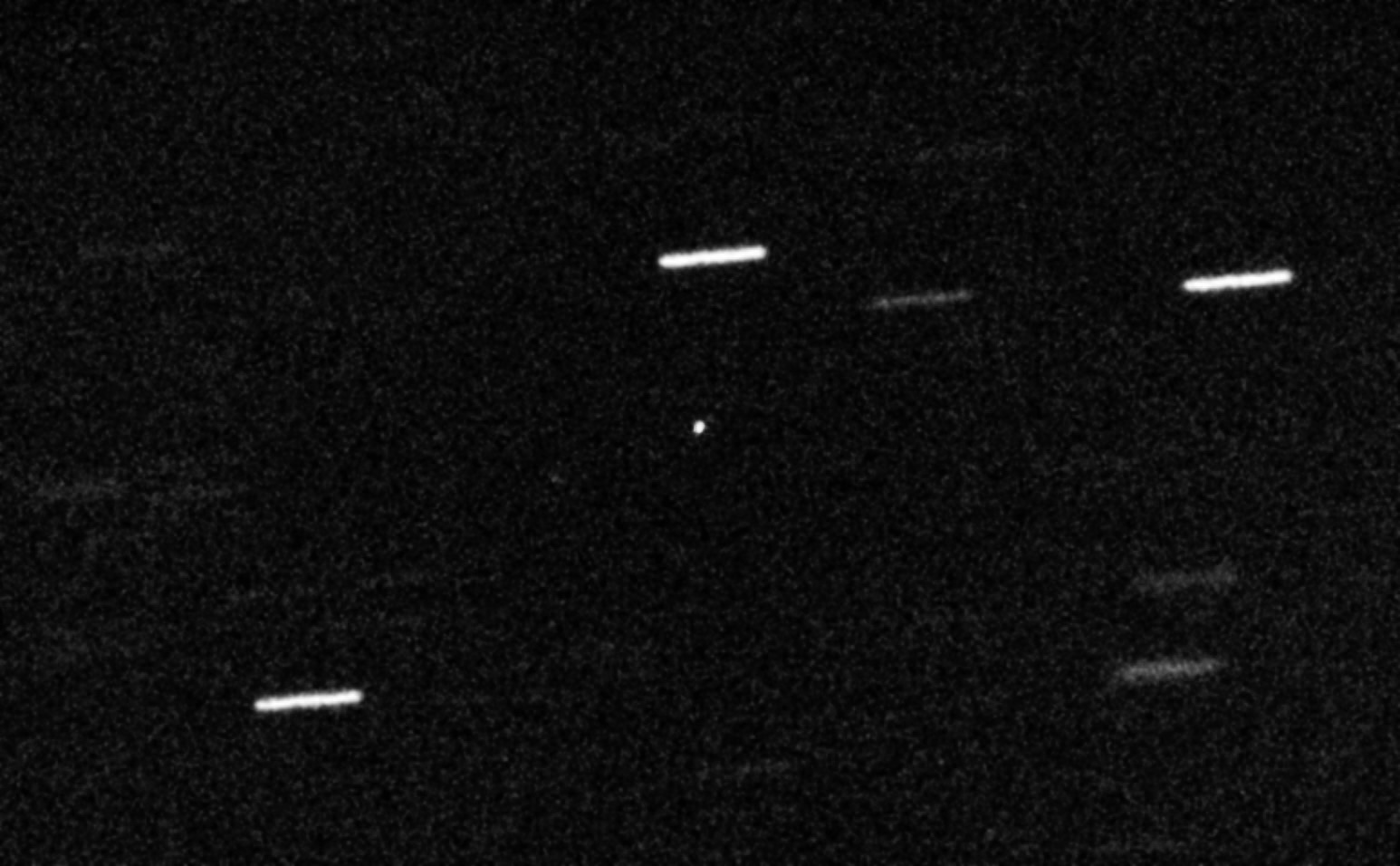}
  \vspace*{-1ex}
  \caption{The central speck of light in this image illustrates the observational challenge associated with identifying passive interstellar probes.  Shown at the center of the image is an object that traversed the Solar System, and the observational challenge is determine whether it is an interstellar comet, i.e., a chunk of rock and ice, or a passive interstellar probe.
    (In this image, the telescope was tracking the object, so the stars appear as streaks.) (Credit: Alan Fitzsimmons [Astrophysics
    Research Centre, Queen's University Belfast], Isaac Newton Group - Instituto de Astrofísica de Canarias)}
  \label{fig:oumuamua}
\end{figure}

Figure~\ref{fig:oumuamua} illustrates the observational challenge.
The possibility of interstellar comets has been recognized for at least two centuries \citep[and references within]{1991ccho.book.....Y,
2022ConPh..63..200S}, and three such interstellar comets now have been recognized \citep{2017Natur.552..378M,2019ApJ...886L..29J,2025ApJ...989L..36S}.
One is traversing the Solar System at the time of writing this paper.
%\citep[e.g.,][]{1975AmJPh..43..590M,1976Icar...27..123S,1982ApJ...255..307V}
\cite{2025arXiv250816825D} summarize the means by which it may be possible to classify an object as a probe rather than an interstellar comet: \emph{anomalous orbit}, \emph{anomalous spectrum}, \emph{anomalous shape}, or \emph{transmissions}.
By definition, a passive probe would produce no transmissions nor respond to any transmissions directed toward it (but an active probe may produce transmissions, \S\ref{sec:matrix.aprobe.comm}).
I now consider the feasibility of using the first three observational approaches for detection and identification of a passive probe.

\subsubsection{Anomalous Trajectory or Orbit}\label{sec:matrix.pprobe.orbit}

Almost by definition, a passive probe's trajectory or orbit likely would not be
anomalous relative to that of an interstellar comet---both would be
dominated by the Sun's gravitational force.  One possibility could be
that a probe's trajectory would exhibit anomalous non-gravitational
accelerations.
For instance, depending upon the size and reflectivity of a probe, its trajectory could be altered by solar radiation pressure.
However, due to outgassing, comets also exhibit non-gravitational accelerations.
Thus, an anomalous trajectory might warrant additional investigation, but, particularly for objects at large distances from the Sun, it might be difficult to reach any definitive conclusions.
(I return to the topic of anomalous trajectories or orbits in considering active probes, \S\ref{sec:matrix.aprobe.orbit}.)

\subsubsection{Anomalous Spectrum}\label{sec:matrix.pprobe.spectrum}

An anomalous spectrum would be one that shows emission or absorption lines or a continuum shape or some combination of these that would not be expected from a natural object.
There is at least one case study illustrating how a spectrum can be used to discriminate between a natural object and a passive probe.

\begin{figure}
  \centering
  \includegraphics[width=0.95\textwidth]
    %alt="Spectra of 2020 SO from 0.75 microns to 2.50 microns.  It is distinguished by a red continuum between 0.75 microns and 1.50 microns, becoming approximately flat longward of 1.50 microns, with absorption features at 1.42 micron, 1.72 microns, and 2.29 microns."]%
  {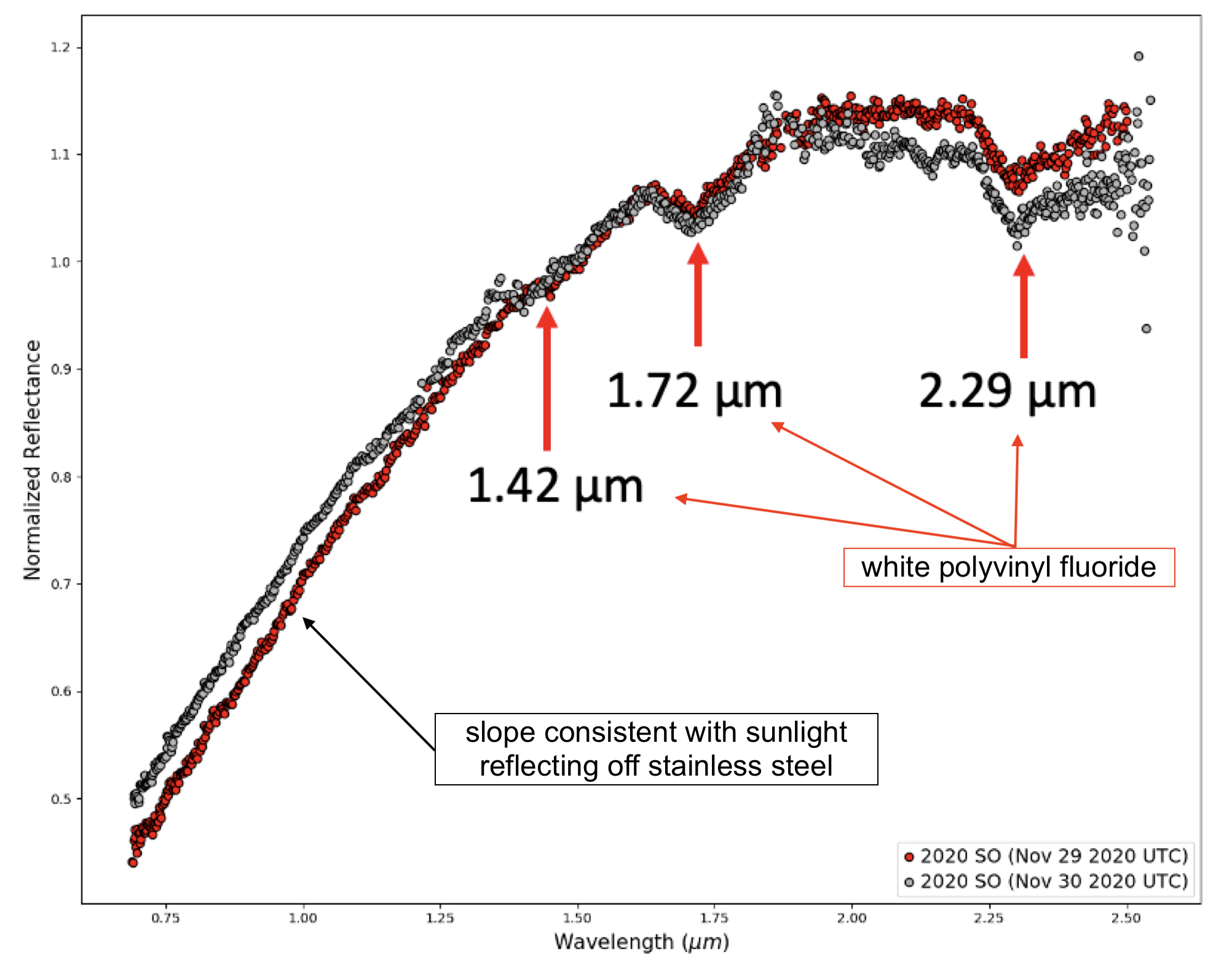}
  \vspace*{-1ex}
  \caption{Using spectra to distinguish a (passive) probe from a natural object.  These near-IR spectra were obtained for the object classified initially as \hbox{2020~SO}.  These spectra show characteristics consistent with the object instead being a Centaur rocket body, leading to its identification as such.
    (Adapted from and reproduced with permission from ``Spectral Characterization of
    \hbox{2020~SO},'' Reddy et al., original copyright \copyright~2001 Advanced Maui Optical and Space Surveillance Technologies
    Conference [AMOS], https://doi.org/10.64861/WYST9178)}
  \label{fig:pprobe.spectrum}  
\end{figure}

The near-Earth asteroid 2020~SO was noted to have a potentially anomalous orbit, moving slowly relative to the Earth.
Motivated by a suggestion that this object might be instead the Centaur rocket body from the Surveyor~2 mission launch, \cite{rbc+21}
obtained near-infrared (near-IR) spectra of the object (Figure~\ref{fig:pprobe.spectrum}).
These spectra contained multiple features consistent with that of a Centaur rocket body.

The combination of an anomalous orbit and anomalous spectra led to the conclusion that 2020~SO is not a near-Earth asteroid, as classified initially, but is more likely a Centaur rocket body.
Key to this conclusion was the combination of multiple measurements
and anomalies---had the orbit of 2020~SO not been anomalous, there might never have been a motivation for obtaining a spectrum of it.

In general, obtaining spectra requires significant investments of telescope time.
Without additional motivation and in light of the large number of (small) Solar System bodies expected to be discovered by the Vera C.~Rubin Observatory's Legacy Survey of Space and Time (\hbox{LSST}, see below), spectra will not be obtained for the vast number of Solar System bodies.

However, an emerging and exciting possibility is that the large number of multi-band photometric observations of small Solar System bodies from the Gaia mission \citep{Gaia,2026arXiv260506838T}; the Spectro-Photometer for the History of the Universe, Epoch of Reionization and Ices Explorer \citep[\hbox{SPHEREx},][]{2022Icar..37114696I}; the \hbox{LSST} \citep{2022FrASS...919168P}; the Euclid mission; and the Near-Earth Object Surveyor Mission \citep[\hbox{NEOSM},][]{NEOSM}
combined with machine learning tools may enable the identification of asteroids with anomalous colors.
There may be some objects with sufficiently anomalous or ``interesting'' colors that would warrant more detailed investigation.

\subsubsection{Anomalous Shape}\label{sec:matrix.pprobe.shape}

For small Solar System bodies, such as asteroids, the classic technique for constraining their shapes has been to interpret changes in brightness as resulting from non-spherical shapes.
Indeed, the dramatic variations in 1I/`Oumuamua's light curve are the basis for the suggestion that it was highly elongated \citep{2017Natur.552..378M}.
%In a similar manner, the significant brightness variations of Saturn's moon Iapetus motivated considerable investigation of its nature \citep{1970Icar...13..282C}.
The obvious confounding factors are that significant brightness variations do not imply an artificial origin nor would a probe necessarily have a shape that would produce significant brightness variations.

Obtaining a resolved image of a probe from ground-based observations is not expected, except under rare circumstances.
For instance, even a probe with a size of approximately 10\,km located at a distance of~1\,au would subtend an angle of only about~0\farcs015, not large enough to resolve with the current capabilities of ground-based visible- and near-IR telescopes.
It has been possible to image some of the larger asteroids, observing their thermal IR emissions with the Atacama Large Millimeter/submillimeter Array \citep[\hbox{ALMA},][]{2015ApJ...808L...2A,2026PSJ.....7...47P}.
Even so, the asteroids imaged have been among the largest in the Solar System, suggesting that a probe would have to have a size approaching 100\,km for this technique to be effective.

\begin{figure}[bt]
 \centering
 \includegraphics[width=0.9\textwidth]
   %alt="Goldstone Solar System Radar images of Asteroid 2024 MK: A sequence of 35 images of 2024 MK acquired on 2024 June 30, in which it appears as a slightly elongated object with highly radar reflective regions rotating in and out of view."]%
  {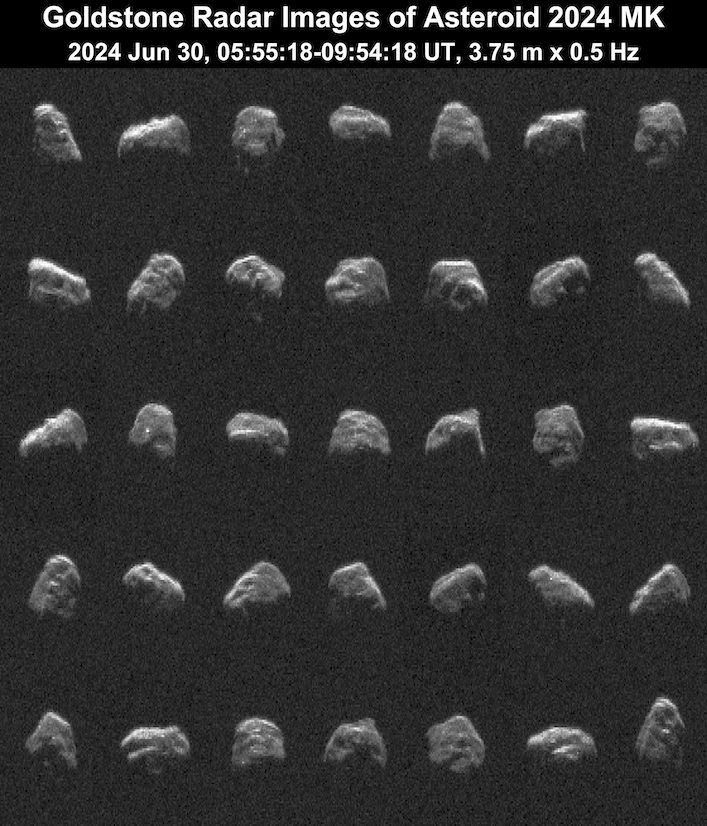}
 \vspace*{-1ex}
 \caption{Illustration of the capability of radar delay-Doppler imaging to provide information about the shape and size of objects in the Solar System.  This delay-Doppler image, obtained with the Goldstone Solar System Radar (GSSR), of the near-Earth asteroid 2024~MK has an effective spatial resolution of~3.75\,m.  While 2024~MK is almost certainly a natural object, radar imaging is likely the only ground-based technique capable of producing resolved images of an interstellar probe, unless the size of the probe were to be substantial.  A significant limitation of radar imaging, however, is that it does require that objects approach reasonably close to the Earth; 2024~MK was at approximately 1~lunar distance when this image was obtained.  (Credit: NASA/JPL-Caltech)}
\label{fig:radar}
\end{figure}

Radar observations with sufficiently high signal-to-noise ratio can be processed to produce so-called delay-Dopper images \citep{2002aste.book..151O}.
In the best cases, radar imaging can obtain information about the shapes and surface features on spatial scales of a few meters (Figure~\ref{fig:radar}).
Such images could provide powerful constraints on the nature of a candidate interstellar probe, with the significant caveat that the candidate probe would have to be quite close to the Earth.
The signal-to-noise ratio for radar observations scales as $1/R^4$, where $R$ is the distance to the object.
Unless the candidate probe passed relatively close to the Earth ($\sim$ lunar distance), its size would have to be substantial for the signal-to-noise ratio to be sufficient for delay-Doppler imaging.

Spacecraft visits to Solar System bodies have provided the highest resolution images of those objects; there have been many suggestions of sending spacecraft to interstellar objects \citep[e.g.,][]{KISSRapidResponse}, most notably `Oumuamua; and the Comet Interceptor mission is designed to provide close reconnaissance of a long-period comet or interstellar object \citep{CometInterceptor}.
A spacecraft flyby or rendezvous with a candidate probe would provide a powerful means of characterizing it.

The production of large numbers (``mass production'') of spacecraft, at least for low-Earth orbit, could introduce the possibility of developing multi-spacecraft reconnaissance missions of the Solar System, and concepts have been developed for reconnaissance of tens to as many as 100 asteroids by a similar number of spacecraft.
However, even if a 100-spacecraft reconnaissance mission of Solar System asteroids were to be financially viable, the number of asteroids and other small bodies in the Solar System is such that the probability of finding a passive interstellar probe by chance would remain quite small.

\subsubsection{Case Study: Detection vs.\ Identification}\label{sec:matrix.pprobe.identify}

As an illustration of the magnitude of the challenge of identifying a passive probe in the Solar System, I consider the following scenario; \cite{2004JBIS...57..283M} has considered a similar scenario.
Suppose that there is a passive probe with a surface area~$A \sim 1\,\mathrm{km}^2$ (i.e., having a diameter of order of~1\,km), perhaps a probe outfitted with a solar sail, that is fully reflective and located at a distance of order 5\,au.
I emphasize that there is no evidence for such a passive probe nor do I have any justification for why there would be such a passive probe.

The apparent visual magnitude of such a passive probe would be $m_V
\sim 19.5$.  The detection of objects with this magnitude is
relatively straightforward.  The challenge is \emph{identification} of
a given 19.5-magnitude object as being a passive probe rather than a
natural Solar System body.
For reference, single LSST observations will probe at least a few magnitudes deeper, and recent projections of the LSST yield are that it will increase the number of Main Belt Asteroids to more than $10^6$, near-Earth objects to more than $10^5$, Jupiter Trojans to more than $10^5$, and even more than $10^4$ trans-Neptunian objects \citep{2025AJ....170...99K}.
Even if there were a large passive probe in
the outer Solar System, determining that it warrants additional
investigation as opposed to more than $10^5$ other objects is the
challenge.

Finally, there also is the consideration that asteroids or comets might be able to be adapted to serve as probes \citep[e.g.,][]{1978QJRAS..19..277P,2026arXiv260316981A}.
Identifying such a probe could be even more challenging than described above.

\subsection{Passive Surface Artifact}\label{sec:matrix.psurface}

\begin{figure}
  \centering
  \includegraphics[width=0.47\textwidth]
    %alt="Region of the lunar surface showing one large crater, a medium-sized crater, many small craters, and reflections from equipment left on the lunar surface by the Apollo astronauts as well as dark tracks from their traverses of the lunar surface."]%
  {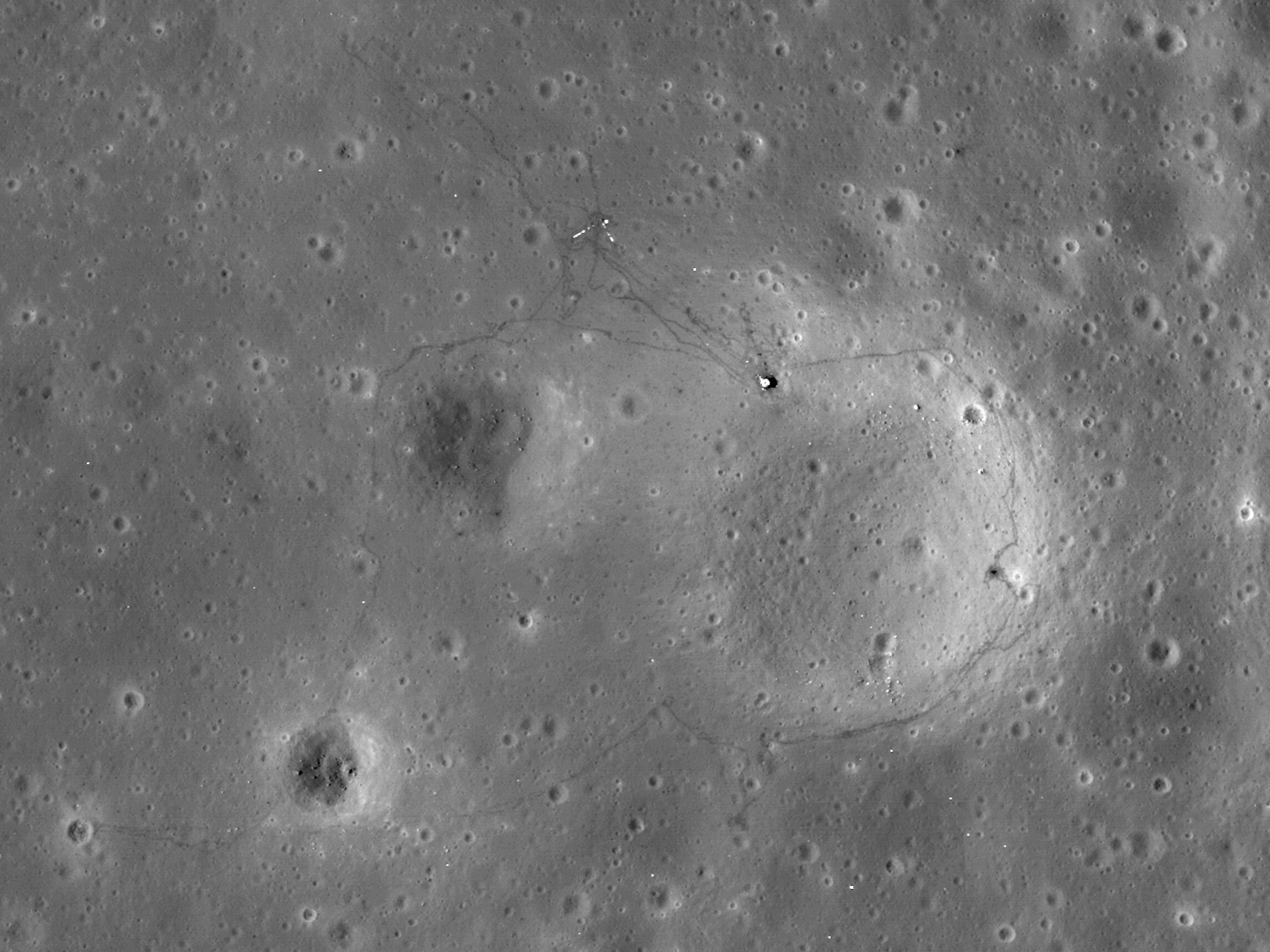}\hfil%
  \includegraphics[width=0.47\textwidth]
    %alt="Region of the Martian surface, appearing as relatively smooth terrain but with many craters also visible.  The location of the Curiosity rover is marked, as are various elements of the system that delivered it to the surface.  In some cases, reflections of those components are visible or local discolorations due to their impacts or use of propulsion also are visible."]%
  {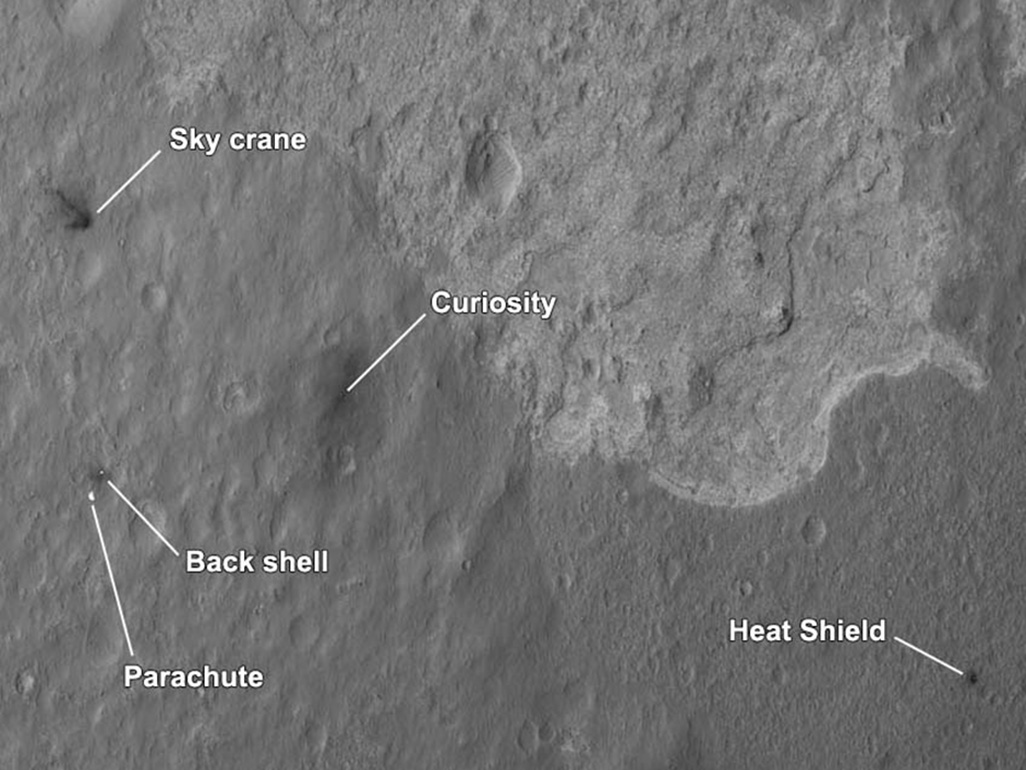}
  \vspace*{-1ex}
  \caption{Illustrations of surface artifacts on two planetary bodies
    in the Solar System from a technological civilization.
    (\textit{left})~Apollo~12 landing site, as imaged by the Narrow
    Angle Camera (NAC) on the Lunar Reconnaissance Orbiter
    (LRO). (\textit{right})~Mars Science Laboratory/Curiosity and
    hardware associated with its delivery to the surface, as imaged by
    the High-Resolution Imaging Science Experiment (HiRISE) camera on
    the Mars Reconnaissance Orbiter (MRO).  (Credit: NASA/GSFC/Arizona
    State Univ.; NASA/JPL-Caltech/Univ.\ of Arizona)}
  \label{fig:surface}
\end{figure}

Figure~\ref{fig:surface} illustrates that there are artifacts on the
surface of planetary bodies in the Solar System that are clear
evidence of a technological civilization, albeit our own civilization.
Nonetheless, that these surface artifacts can be identified from
orbit is an existence proof of such artifacts and suggests that a
search for additional such artifacts is warranted
\citep{2013AcAau..89..261D}.

In principle, this search technique is straightforward: Search all imagery of planetary surfaces from international space agencies (\hbox{NASA}, European Space Agency [ESA], Japan Aerospace Exploration Agency [JAXA], \ldots) for indications of anomalous structures.
In practice, there is a significant numerical challenge.

Consider just the surface of the Moon.
The Narrow Angle Camera (NAC) on the Lunar Reconnaissance Orbiter (LRO) can obtain a surface resolution of approximately 0.5\,m \citep{2016SSRv..200..357S}.
The Moon has a radius of approximately 1700\,km.
Thus, for the Moon alone, there could be as many as $72 \times 10^{12}$~pixels to examine.
Moreover, this search should be conducted in a reliable and reproducible manner.

Machine learning tools offer the possibility for reducing substantially the fraction of planetary imagery that need to be examined, thereby advancing this technosignature search.
There have been initial efforts to conduct such searches (\citealt{2024IJSTA..17.6589L}, see also an example in \citealt{KISS}), but there also are considerable opportunities to expand these initial efforts.

There are several cautionary elements related to searching for surface
artifacts.  Table~\ref{tab:surfaces} provides an illustration of the
completeness of surveys of Solar System objects with solid surfaces.
There are many objects, mostly the small bodies in the Solar System,
that never have been imaged.
Even for those objects that have been imaged, typically only a fraction of the surface has been imaged.
Further, in most cases, the imagery is relatively coarse, with surface resolutions of order 1\,km common.
The consequence is that substantial technological objects could remain unrecognized.
As a specific example, a structure 1\,km in size would occupy less than one pixel for essentially all of the surfaces of Saturn's moons and likely would go unnoticed.

\begin{table}
  \centering
  \caption{Completeness of Surveying Solar System Bodies\label{tab:surfaces}}
  \begin{tabular}{lcl}
  \noalign{\hrule\hrule}
  \textbf{ } & \textbf{Best Case} & \textbf{ }\\
  \textbf{Object} & \textbf{Spatial Resolution} & \textbf{Source} \textbf{[Reference]}\\
  \noalign{\hrule\hrule}
  Mercury              & 250\,m                            & MESSENGER~[1] \\
  Venus	               & 120\,m--300\,m                    & Magellan~[2] \\
  Moon	               & $\sim 0.5\,\mathrm{m}$ over fraction of surface & Lunar Reconnaissance \\
                       & 100\,m for global coverage                      & \qquad{}Orbiter~[3,4] \\
  Mars	               & $\sim 1\,\mathrm{m}$ over about~1\% of surface & Mars Reconnaissance Orbiter~[5] \\
                       & 100\,m for global coverage                     & \textit{The Atlas of Mars}~[6] \\
  Main Belt Asteroids  & $\approx 0.5\,\mathrm{km}$                   & Ceres: Dawn~[7] \\
  Jovian satellites    & $\sim 10\,\mathrm{m}$             & Ganymede, Europa: Galileo~[8] \\
  Saturnian satellites & $\lesssim 1\,\mathrm{km}$             & Iapetus: Cassini~[9] \\
                       &                                   & cf.\ \textit{2001: A Space Odyssey} \\
  Uranian satellites    & $\gtrsim 10\,\mathrm{km}$ over fraction of surfaces & Voyager~2 flyby~[10] \\
  Neptune/Triton       & 3\,km over 1/3 of surface         & Voyager~2 flyby~[11] \\
\noalign{\hrule\hrule}
\end{tabular}
\parbox{0.95\textwidth}{%
    {\footnotesize
    Spatial resolution values listed should be taken as representative.
    References:
    [1]~\cite{2018SSRv..214....2D}; 
    [2]~\cite{1991Sci...252..260P};
    [3]~\cite{2012JGRE..117.0H17S}, [4]~\cite{2016SSRv..200..357S};
    [5]~\cite{MRO};
    [6]~\cite{MarsAtlas};
    [7]~\cite{2016P&SS..121..115R};
    [8]~\cite{2000AdSpR..26.1641B};
    [9]~\cite{2008P&SS...56..109R};
    [10]~\cite{1986Sci...233...43S};
    [11]~\cite{1989Sci...246.1422S}; 
    Images and data from much of the spacecraft reconnaissance of the Solar System are available from NASA's Planetary Data System (\hbox{PDS}, https://pds.nasa.gov) and the European Space Agency's Planetary Science Archive (\hbox{PSA}, https://psa.esa.int/).
    }}
\end{table}

A key aspect for passive surface artifacts is that, even if passive today, they must have been active at one time in the past.
As a simple example, an object with about the mass of a CubeSat ($\sim 10\,\mathrm{kg}$) could enter the Solar System with a velocity of order 50\,km\,s${}^{-1}$.  
Its kinetic energy would be of order \hbox{13\,GJ} ($\sim 3\,\mathrm{tonne}$ \hbox{TNT}).
If it has no means of slowing its approach to a planetary body, it simply will produce a crater.  
Indeed, there has been active discussion in the literature as to whether some of the air blasts detected by U.{}S.\ Government sensors represent explosions of small interstellar objects in the Earth's atmosphere
%\textbf{CNEOS 2014-01-08 Peña-Asensio, Trigo-Rodríguez, & Rimola}
\citep{2019arXiv190407224S,2019RNAAS...3...68Z,2022AJ....164...76P}.

Even if an interstellar probe produced a crater on the surface of a planetary body, one might speculate about whether there could be
sufficient abundance anomalies in or near the crater to identify it as the remnant of a technological artifact.
Such an approach would be analogous to how an anomalous abundance of iridium at the K-Pg boundary
between the Cretaceous and Paleogene epochs supports the idea that a comet impacted the Earth approximately 65\,Myr ago.
There are clear challenges in using abundance anomalies to identify potential impact sites of interstellar probes.  Multiple uncertainties
would have to be addressed: Would the abundance anomalies be
sufficiently widespread or notable that they could be detected from
orbit or would \textit{in situ} sampling be required?  What kind of
elements or compounds should be considered?  If an anomaly is
detected, is it sufficient to conclude that an interstellar probe
impacted or are there confounding factors?

A final cautionary note about surface artifacts is their persistence.
\cite{2018IJAsB..17...96W} has assessed the likelihood that passive surface artifacts both would survive and be recognizable on surfaces in the Solar System.
Various processes are operative on different Solar System bodies, from resurfacing to micrometerorite impacts to aeolian activity (at least on Mars), but the overall result is that recognizing passive surface artifacts as ET technological objects might be quite difficult if they are much older than 1\,Myr.

\subsection{Active Probe}\label{sec:matrix.aprobe}

If there were an active probe in the Solar System, there are multiple ways that it could generate detectable signatures, including its
propulsion system, its communication system, its ``waste heat,'' or its trajectory or orbit. 
I consider these possible avenues for detection and identification of an active probe, which correspond roughly to considering different wavelengths at which signatures might occur.

\subsubsection{Propulsion System}\label{sec:matrix.aprope.propel}

A probe might generate a detectable signature at X- or $\gamma$-ray wavelengths, or by emitting high-energy particles, if it had a nuclear power or propulsion system \citep{1994OLEB...24..344P}.
There has been a long history of using nuclear power systems for deep space missions, such as Voyager, Galileo, Cassini, and New Horizons, and and many investigations of nuclear propulsion systems \citep{AtomicPowerII,THOMAS2024362}.
Indeed, as this paper was being written, the NASA Administrator announced a renewed focus on nuclear propulsion systems for deep space spacecraft.

However, in general, one might expect that any nuclear system would be
heavily shielded.  In typical designs of nuclear power systems, the
high-energy photon or particle products of the nuclear reactions are
converted to another form of energy for use.  For instance, in a
radioisotope thermoelectric generator (RTG), the heat generated by the
nuclear reactions is converted to electricity.
Thus, there is a strong incentive for as few decay products or as little heat as possible to escape so as to obtain a high power efficiency.

Similarly, in typical
designs for nuclear propulsion systems, a nuclear reactor is used to
heat a propellant, which then is expelled at high velocity from a
rocket engine.  Thus, if there are significant high-energy emissions
from the nuclear propulsion system, then it likely is
inefficient.

Finally, if there are significant emissions from the nuclear propulsion or power system, there could be the risk that other systems of the interstellar probe would be damaged.
Thus, it seems difficult to motivate or justify a search for significant sources of X- or $\gamma$-ray emission, or high-energy particles, from within the Solar System because it is not clear what meaningful constraints could be obtained if no detection is made, particularly in contrast to searches at other wavelengths.
(See below.)

More generally, the visible or IR spectrum of the exhaust of terrestrial propulsion systems shows emission lines from the compounds or atoms used in the systems, though there tends not to be extensive descriptions of the spectral signatures in the open literature \citep[e.g.,][]{s00}.
Nonetheless, much like the discussion of passive probes (\S\ref{sec:matrix.pprobe.spectrum}), if there is a motivation to obtain a spectrum of an ``interesting'' object, anomalous emission lines may be a signature of the propulsion system of an active probe.

\subsubsection{Waste Heat}\label{sec:matrix.aprobe.heat}

A probe should have an equilibrium temperature due to its exposure to
sunlight, and, by Planck's law, it should produce a spectrum of radiation.
Within the Solar System, a probe's spectrum should peak in the \hbox{IR}.
Thus, a search technique would be to look for an object having a spectrum that deviates significantly from that expected given its distance from the Sun.
Indeed, while there is a strong incentive for the power and other systems on a probe to be efficient, they cannot be 100\% efficient.
Heat will result as a by-product of these various activities, leading to the expectation that a probe would generate so-called ``waste heat,'' and suggesting that an active probe would appear to be ``too hot'' given its distance from the Sun.

Should a probe have a chemical propulsion system, its exhaust would have some temperature, which likely also would result in it appearing anomalously hot. 
\cite{2020arXiv200704892H} also noted that, should a probe be traveling at sub-relativistic speeds ($\sim 0.1c$) through the Solar System, it would radiate simply due to the heat resulting from dust impacts.

Intriguingly, observations of asteroids by the Wide-field Infrared
Survey Explorer (WISE) have revealed multiple objects having spectra
inconsistent with their expected temperatures
\citep{2011ApJ...737L...9M}.  However, based on these
observations alone, it is not possible to conclude that any of these
objects represent active interstellar probes.  Developing a model for
the temperature and resulting thermal emission of an asteroid involves
several parameters, not all of which are known with high certainty.
Further, very often assumptions or simplifications must be made.

Nonetheless, finding asteroids (or comets) with unusual physical properties could be rewarding, even if no interstellar probes are found.
SPHEREx \citep{2022Icar..37114696I} is conducting and \hbox{NEOSM} \citep{NEOSM} will conduct observations at IR wavelengths and could reveal many objects warranting additional investigation.

\subsubsection{Communication Systems}\label{sec:matrix.aprobe.comm}

There is extensive discussion in the literature concerning the detection of transmissions from active probes, either unintentionally or transmissions that would be aimed at Earth.  
Based on terrestrial practices, discussions typically focus on transmissions either at radio- or at visible--near-IR wavelengths (i.e., lasers).

I begin by considering a probe with a Voyager-like telecommunication system \citep{VoyagerTelecomm}, namely one that transmits at a wavelength of 3.6\,cm (8420\,MHz, ``X~band'') using a 3.66\,m antenna.
The resulting antenna beam or point spread function is approximately $0.\!\!^\circ5$, or the subtended solid angle is $2.4 \times 10^{-4}\,\mathrm{sr}$.
Therefore, at any given instant, the probe's transmissions are focussed into a region that is approximately 0.002\% of the entire sky, i.e., the probability of intercepting the probe's transmissions is 0.002\%.

One might wonder if a Voyager-like telecommunication system would be sufficient to produce detectable signals (or ``close the link'') over interstellar distances.\footnote{
``Six Deep Space Network Antennas in Madrid Arrayed For the First Time,''
\texttt{https://www.jpl.nasa.gov/images/pia26147-six-deep-space-network-antennas-in- madrid-arrayed-for-the-first-time/}}
Unfortunately, most techniques for increasing the performance of a telecommunication system make it more difficult to detect.
One means of obtaining a higher-performance telecommunication system would be to increase the diameter~$d$ of the antenna carried by the probe.\footnote{
A larger antenna on a probe would tend to reflect more sunlight, making it potentially more detectable.  
Nonetheless, as discussed in Section~\ref{sec:matrix.pprobe.identify}, it likely still would be difficult to identify as a probe.}
The solid angle subtended, and therefore the probability of intercept, scales as $d^{-2}$.  
Thus, considering even a modestly larger antenna, e.g., 12\,m such as has been used for some Earth-orbiting telecommunications satellites, would decrease the probability of intercept by approximately $10\times$.

Even if the probability of intercepting (or detecting) a probe's transmissions might be small, there would be a reasonable chance that it could be identified as being a technosignature signal. 
A probe and the Earth would be moving relative to each other due to their orbits in or through the Solar System.
Taking a Voyager-like probe to be about~1\,au away and a characteristic relative velocity to be about~30\,km\,s${}^{-1}$, comparable to the Earth's orbital velocity, the probe's transmissions would remain directed toward the Earth for a time scale of about~10\,hr.
If a probe were in the field of view of a radio interferometer, the probe would appear as an unresolved, narrowband, slowly-moving transient, moving across about~$0.\!\!^\circ1$ to~$0.\!\!^\circ5$ in a few to several hours.
There are astronomical radio transients that evolve on time scales of tens of hours \citep[e.g.,][]{1999ApJ...522L..97K,2006ApJ...650..261S}, but they are broadband sources that are sufficiently distant so as to appear stationary.
A narrowband point source moving across the sky would be unlike any known celestial source, and its proper motion across the sky would be unlike those of satellites in Earth orbit.
Crucially, the duration that the probe would be in the radio interferometer's field of view likely would enable collecting sufficient data to identify the received signal as a technosignature.
Further, depending upon the frequency and the size of the radio interferometer, it is possible that a probe would be in the far-field of the interferometer while most or all satellites would be in the near-field, providing another means of discriminating against Earth-orbiting satellites.
I stress that the objective of interferometric imaging would not be to resolve the probe itself, but to  discriminate it from terrestrial and near-Earth transmitters \citep{2005RaSc...40.5S18H,2012AJ....144...38R,KISS,2025AJ....169..122T,khh+26}.

A laser communications system for deep space, operating at~1.5\,$\mu$m
(in the near-IR), has been demonstrated on NASA's Psyche Discovery
mission \citep{DSOC}.  The solid angle subtended, and probability of
intercept, scales as $\lambda$.  Thus, if interstellar probes are
using laser telecommunication systems, the probability of intercepting
their transmissions would be a factor of~$10^4$ or more \emph{smaller} than detecting radio transmissions, and their transmissions would remain aligned with the Earth for only a few seconds.
Indeed, one of the motivations for adopting laser telecommunications systems is precisely the low probability of intercept, which results in higher security for the transmitted signals.

%Only if the probe is transmitting toward Earth intentionally, compensating its pointing so as to keep the transmission beam aligned on the Earth, would it be feasible both to detect and verify a signal.

\subsubsection{Anomalous Trajectory or Orbit}\label{sec:matrix.aprobe.orbit}

Even if the exhaust or waste heat from an active probe's propulsion system would be difficult to detect, it could result in an anomalous trajectory or orbit.  
There are two general manners in which a probe's trajectory could appear anomalous relative to other Solar System bodies, particularly asteroids: it could exhibit \emph{discontinuous} or \emph{secular} changes.

A dramatic case study of a discontinuous change in an object's orbit was that of 2001~DO${}_{47}$ \citep{2001IAUC.7589....3G}.
Identified initially as a near-Earth asteroid, during the course of multiple observations over the course of one week with the Goldstone Solar System Radar (GSSR), its trajectory was observed to have changed in a discontinuous manner.
As a result of this change in its orbit, the object was confirmed to be the Wind spacecraft, which had fired its engines between various GSSR observations.
In one sense, the case of 2001~DO${}_{47}$ represents the identification of an active probe in the Solar System, albeit a probe from our civilization rather than another.

Less dramatic, though equally relevant case studies involve so-called ``erratic comets,'' comets with multi-decadal observational spans that show sudden, potentially discontinuous changes in their orbits \citep{1971AJ.....76.1135M}.
Notably, many or all of these orbital changes did not occur following or near to a close approach to Jupiter, thereby excluding any gravitationally-induced changes, including tidal effects, on the comets.
It is not clear that any single process can explain the orbital changes of erratic comets, though \cite{1971AJ.....76.1135M} favor an explanation in which the comets' nuclei are impacted by a smaller body, which likely would result in rocket-like expulsion of sub-surface material from the impact itself or subsequent sublimation of exposed material or both.

The clear challenge in using discontinuous orbital changes to identify an active probe is having a sufficient number of observations.
If the number of observations is too small, a far more likely possibility is that a probe would be mistaken for two (or more) asteroids on different orbits, which could be considered to be a subset of the more general ``asteroid attribution'' problem \citep{2001Icar..151..150M}.

A secular change in an object's orbit is one in which the discrepancy between an object's predicted and observed orbits increase with time. 
A secular orbital change would result from one or more non-gravitational forces acting on the object.
In the context of an active probe, a secular orbital change could occur if its shape resulted in considerable solar radiation pressure or a propulsion system that produced a small but continuous thrust, such as solar electric propulsion (SEP).
A confounding factor for using a secular orbital change to identify an active probe is that asteroids and comets also exhibit secular orbital changes due to non-gravitational forces. 
Perhaps the most well-known such non-gravitational forces are volatile outgassing and the Yarkovsky effect.
Volatile outgassing, particularly if from localized regions on a small body's surface \citep{2015Sci...347a0709G,2019A&A...630A...4K}, not only produces the comae for which comets are known but also can produce an effective thrust.
The Yarkovsky effect, resulting from the thermal inertia of an asteroid, also results in an asymmetric effective thrust.
Much like the case for discontinuous orbital changes, secular orbital changes also can result in ``asteroid attribution'' problems, if the non-gravitational force is sufficiently large but the object is observed over only a short interval.

Intriguingly, a class of asteroids termed ``dark comets'' has been identified recently from orbital changes consistent with non-gravitational forces, but displaying no comae or other effects of outgassing and, in some cases, inconsistent with the Yarkovsky effect \citep{2023PSJ.....4...35S,2023PSJ.....4...29F}.
There has been discussion in the literature regarding whether one of these dark comets might be a defunct Venera spacecraft \citep{2025arXiv250303552L,2025RNAAS...9...55M,2025RNAAS...9...58S,2025arXiv250309668L,2025arXiv250609478H}.
While dark comets likely are asteroids, their orbital changes also are consistent with those that might result from active probes conducting occasional orbital maneuvers or having low-thrust propulsion systems \citep{2025arXiv250816825D,DarkCometProbes}.
In particular, the accelerations (or non-gravitational forces) required to explain the orbits of dark comets are well within the achieved capabilities of spacecraft propulsion systems that our civilization has produced.

While an anomalous trajectory or orbit likely would not be sufficient to identify an active probe, as the example of 2020~SO demonstrated, an anomalous orbit may be the first clue that an object is an active probe.  
Conversely, these examples also illustrate how an active probe might be classified initially as an asteroid.
New approaches, often involving elements of machine learning, are being developed to find asteroids in data from surveys such as the Zwicky Transient Facility (ZTF), Transiting Exoplanet Survey Satellite (TESS), and the \hbox{LSST} \citep{2019PASP..131c8002M,2026arXiv260512391P}.
These and other future observations and analyses should include the possibility that probes would be among the objects detected.

\subsection{Active Surface Artifact}\label{sec:matrix.asurface}

Detecting and identifying an extraterrestrial technology that is active on the surface of a planetary body, or perhaps more generally active in a planetary atmosphere, is somewhat the union of the considerations for passive surface artifacts (\S\ref{sec:matrix.psurface}) and active probes (\S\ref{sec:matrix.aprobe}):  There is a significant fraction of the Solar System that has never been surveyed or, at best, only coarsely surveyed and communication activities likely have a low probability of being intercepted.
Thus, it is possible that there is one or more active objects on the surface or surfaces of the Solar System bodies that have gone unnoticed.

One potentially promising avenue would be to consider searches for anomalous signatures at visible or IR wavelengths, particularly among asteroids.
There have been many suggestions that asteroids in the Main Belt would present ready sites for resources and that exploitation of those resources (``mining'') would produce detectable signatures \citep{1978QJRAS..19..277P,1983AcAau..10..709P,1998JBIS...51..175K}.
%\cite{2011IJAsB..10..307F} have extended this concept to consider the possibility of detecting signatures of mining of asteroids in other planetary systems.
%One potential signature would be increased IR emission due to the dust generated.
If there were activity on an asteroid, it could generate dust or cause other loss of mass, which could result in the asteroid having anomalous properties at the visible or IR wavelengths, due to increased scattering of sunlight or enhanced thermal emission or both.
In this respect, the discovery of a class of so-called ``active asteroids'' \citep{j12,bdm15} is an example of how such active surface artifacts might be recognized.
Active asteroids are objects showing properties consistent with asteroids at discovery, but later revealing at least some properties consistent with comets, such as outgassing.
Similar to the discussion about asteroids with anomalous IR signatures (\S\ref{sec:matrix.aprobe}), most likely, all active asteroids are simply asteroids experiencing transient mass loss by some natural process or processes.

Much like the case for an active probe, an active surface artifact also should generate waste heat.  
In comparison to a passive surface artifact, searching for wast heat does not require that the object resolved---a single resolution element (pixel) that is at an anomalous temperature could be a marker for an active surface artifact.
Thus the angular resolution requirements for a thermal IR imager can be relaxed relative to those of an imager that would need to resolve an artifact.
Even if no active surface artifacts are found, a thermal IR survey of solid body surfaces could yield other discoveries such as previously-unrecognized sites of vulcanism or recent impact sites.
The detection of regions of active vulcanism on Io \citep{2025JGRE..13008850P} and of enhanced thermal emission at the site of the Lunar Crater Observation and Sensing Satellite (LCROSS) impact with the LRO/Diviner radiometer \citep{2010Sci...330..477H} both represent illustrations of how  such surveys could be conducted.

%Nonetheless, even if additional investigation of active asteroids or other bodies showing anomalous properties revealed no active extraterrestrial technologies, improved understanding of Solar System processes certainly would result.

The challenge for conducting such a survey, particularly in comparison with searching for waste heat from an active probe, would be the required measurement precision as a result of ``beam dilution.''  
Whereas an active probe would be seen against a cold, deep space background ($\approx 3\,\mathrm{K}$), an active surface artifact would be sitting on a much hotter background.
The extent to which it could be distinguished from that background depends upon the relative ratios of its temperature and area to that of its surroundings and the pixel scale.
As a specific example, suppose that there were to be a mining facility located on an asteroid.
For the sake of illustration, I consider a facility that is of order 30\,m in size ($A_{\mathrm{factory}} \sim 900\,\mathrm{m}^2$) at a uniform temperature of~1000\,\hbox{K}.
Suppose that the surface of the asteroid has a characteristic temperature of order 100\,\hbox{K} and that it is surveyed with a pixel scale of order 1\,km (viz.\ Table~\ref{tab:surfaces}).
The relative contrast of the mining facility to its surroundings is then $(1000\,\mathrm{K}/100\,\mathrm{K})(900\,\mathrm{m}^2/10^6\,\mathrm{m}^2) \approx 0.01 \approx 1\%$.
The LRO/Diviner radiometer did achieve comparable precisions at some of its bands \citep{Diviner}, and it may be possible to achieve better precision by conducting relative measurements.

Thus, searching for waste heat on planetary surfaces is feasible, but there could be confounding natural phenomena that could make it difficult to identify an active ET technology.

\section{Concluding Thoughts and Next Steps}\label{sec:conclude}

Returning to the original hypothesis---that there might be ET technological objects in the Solar System---it is clear that I have failed to falsity it.
Indeed, it is quite possible that multiple probes or surface objects could be present without being recognized.

A recurring theme in this assessment has been the difficulty in identifying an ET technological object.
There have been multiple instances in which technological objects from our civilization have been mistaken for natural objects.
Even if an ET technological object is detected, there likely will be a host of natural objects or processes that will serve as confounding factors.
However, in the spirit of Dyson's First Law for Searches for Extraterrestrial Intelligence,\footnote{
Every search for alien civilizations should be planned to give interesting results even when no aliens are discovered. 
}
these confounding factors also can serve as part of the motivation for conducting surveys for ET technological objects---even if no ET technological objects are found, a better census of the Solar System, and a better understanding of the processes operating within it, will result.

This assessment has not considered the probability of success, arguing only that humanity's own interstellar probes provide an existence proof for the possibility of ET interstellar probes.
Much like the case for discussions of technosignatures more generally, predictions in the literature range from highly pessimistic to optimistic.
Perhaps the classic series of exchanges on this topic, reflecting the diversity of opinions, was that between Hart \& Zuckerman and Sagan \& Newman \citep{1975QJRAS..16..128H,1981Icar...46..293N,1982ewat.book.....H,1982ewat.book....1H,1983QJRAS..24..113S}.

Thus far, I have approached the question of signatures of
extraterrestrial technology in the Solar System from the scientific
and technical perspectives---given current knowledge about the Solar System and current practices and attempting to
make plausible extrapolations, how likely would it be to detect any such technosignatures. 
A separate question, more programmatic (a.{}k.{}a.\ political) in nature is, What if a technosignature is detected in the Solar System?

In many people's minds, there likely will be a substantial difference between a technosignature from a distant planet, star, or galaxy\footnote{
Once upon a time, the author was attempting to convey the excitement about the discovery and study of extrasolar planets.  At the conclusion of the discussion, his grandmother stated, ``Yes, but they are so far away.''}
and something that is ``Here! Now!''
Most of the focus for programmatic discussions about technosignatures historically has been in the context of received signals \citep[e.g.,][]{2025arXiv251014506G}, rather than physical artifacts, though \cite{2012AcAau..78...31B} and \cite{2020SpPol..5201377W} have commented on the possible consequences of and actions resulting from the discovery of physical artifacts.
(See also \citealt{2026arXiv260501019B}.)

A significant development in the past decade has been the rise of capability in entities other than nation-states to conduct space missions.
While much of this capability has been focussed on low-Earth orbit (LEO), there are at least expressions of interest in moving beyond \hbox{LEO}, and initial demonstrations of private missions to the Moon, albeit often heavily government-subsidized.
If a physical artifact were to be found in the Solar System, depending upon where it were to be located and its nature, there could be non-governmental entities interested and even capable of accessing or retrieving it.
In addition to the potential for financial return, or perception of potential financial return, non-governmental entities might be motivated by other considerations.
For instance, there could be value perceived simply by acquiring an artifact, in a manner similar to how some private collectors acquire items in part because of the interest in doing so.

These concerns notwithstanding, a thesis of this paper is there is considerable discovery space remaining within the Solar System and that effective searches could be conducted with existing data sets or those to be acquired in the next few decade.
Even if no ET physical technosignature is identified, investigating anomalies in those data sets likely will produce discoveries in astronomy and planetary science.

%%%%%%%%%%%%%%%%%%%%%%%%%%%%%%%%%%%%%%%%%%%%%%%%%%%%%%%%%%%%%%%%
% Acknowledgements
\bigskip 

I thank K.~Denning, J.~Giorgini, G.~Hellbourg, M.~Hofstadter, L.~Johnson, J.~Kasper, A.~Mahabal, J.~Masiero, R.~Park, U.~Rebbapragada, D.~Seligman, K.~Wagstaff, J.~Weiss, and J.~Wright for helpful and illuminating conversations on many of the topics discussed here.
I acknowledge ideas and advice from the participants in the Data-Driven Approaches to Searches for the Technosignatures of Advanced Civilizations workshop organized by the W.~M.~Keck Institute for Space Studies.
This research has made use of NASA's Astrophysics Data System.
Part of this work was carried out while the author was at the Jet Propulsion Laboratory, California Institute of Technology, under a contract with the National Aeronautics and Space Administration.

%%%%%%%%%%%%%%%%%%%%%%%%%%%%%%%%%%%%%%%%%%%%%%%%%%%%%%%%%%%%%%%%
% Bibliography
\clearpage
%\bibliographystyle{iaulike}
%\bibliography{SolarSystemTechnosignatures}

\end{document}